\documentclass{PoS}

\newcommand{\be}{\begin{eqnarray}}
\newcommand{\ee}{\end{eqnarray}}
\newcommand{\nee}{\nonumber\end{eqnarray}}

\newcommand{\drbar}{{\overline{\rm DR}}}

\newcommand{\mch}[1] {m_{\ti \x^+_{#1}}}
\newcommand{\mnt}[1] {m_{\ti \x^0_{#1}}}

\newcommand{\msg}    {m_{\ti g}}
\newcommand{\msu}[1] {m_{\ti u_{#1}}}

\def\gev             {{\rm GeV}}

\def\be            {\begin{equation}}
\def\ee            {\end{equation}}
\def\bea            {\begin{eqnarray}}
\def\eea            {\end{eqnarray}}

\def\a              {\alpha}
\def\b               {\beta}
\def\d               {\delta}

\def\x               {\chi}
\def\ti              {\tilde}
\def\sq              {\ti q}

\def\su                {\ti{u}}
\def\sto                  {\ti{t}}
\def \sca                 {\ti{c}}

\def\dll            {\d^{LL}_{23}}
\def\durr            {\d^{uRR}_{23}}

\def\dulr            {\d^{uLR}_{23}}

\title{$h^0(125GeV) \to c \bar{c}$ as a test case for quark flavor violation in the MSSM}

\ShortTitle{$h^0(125GeV) \to c \bar{c}$ as a test case for quark flavor violation in the MSSM}

\author{\speaker{Keisho Hidaka}\\
        Department of Physics, Tokyo Gakugei University, Tokyo 184-8501, Japan\\
        E-mail: \email{hidaka@u-gakugei.ac.jp}}

\author{A. Bartl\\
        Universit\"at Wien, Fakult\"at f\"ur Physik, A-1090 Vienna, Austria\\
        E-mail: \email{Alfred.bartl@univie.ac.at}}

\author{H. Eberl, E. Ginina and W. Majerotto\\
        Institut f\"ur Hochenergiephysik der \"Osterreichischen Akademie der Wissenschaften, A-1050 Vienna, Austria\\
        E-mail: \email{helmut.eberl@oeaw.ac.at, elena.ginina@oeaw.ac.at, and Walter.Majerotto@oeaw.ac.at}}

\abstract{
We calculate the decay width of $h^0(125GeV) \to c \bar{c}$ in the Minimal Supersymmetric 
Standard Model (MSSM) with \emph{non-minimal} quark flavor violation (QFV) at full one-loop level. 
We adopt the $\overline{\rm DR}$ renormalization scheme.
We study the effects of the mixing of the second and third squark generations (i.e. scharm-stop mixing)
on the decay width, respecting the experimental constraints from B-meson data, the Higgs mass measurement 
and supersymmetric (SUSY) particle searches. 
We show that the decay width $\Gamma (h^0 \to c \bar{c})$ at the full one-loop level 
is very sensitive to the SUSY QFV parameters.
In a scenario with large scharm-stop mixing, the decay width can differ up to 
$\sim \pm 35\%$ from its SM prediction.
After taking into account the experimental and theoretical uncertainties of the decay width, 
we conclude that these QFV SUSY effects can be observed at a future $e^+ e^-$ 
collider such as ILC (International Linear Collider).
}

\FullConference{The European Physical Society Conference on High Energy Physics\\
        22-29 July 2015\\
        Vienna, Austria}

\begin{document}

\section{Introduction}

The properties of the Higgs boson with a mass of 125 GeV, discovered 
by ATLAS~\cite{Higgs@ATLAS} and CMS~\cite{Higgs@CMS} at the LHC (Large Hadron Collider), are
consistent with the predictions of the Standard Model (SM). 
It is the most important issue in the present particle physics world to determine 
if it is the SM Higgs boson or a Higgs boson of 'New Physics'. 
In this article based on our paper \cite{Bartl}, 
we study the possibility that it is the lightest Higgs boson $h^0$ of the Minimal Supersymmetric 
Standard Model (MSSM), by focusing on the decay $h^0 \to c \bar{c}$, where c is a charm quark. 
We calculate the decay width $\Gamma(h^0 \to c \bar{c})$ at full one-loop level in the $\drbar$ 
renormalization scheme in the MSSM with \emph{non-minimal} quark flavor violation (QFV).

\section{Definition of the QFV parameters}

In the super-CKM basis of $\sq_{0 \gamma} =
(\sq_{1 {\rm L}}, \sq_{2 {\rm L}}, \sq_{3 {\rm L}}$,
$\sq_{1 {\rm R}}, \sq_{2 {\rm R}}, \sq_{3 {\rm R}}),~\gamma = 1,...6,$  
with $(q_1, q_2, q_3)=(u, c, t),$ $(d, s, b)$, one can write the squark mass matrices in their most general $3\times3$-block form
\begin{equation}
    {\cal M}^2_{\tilde{q}} = \left( \begin{array}{cc}
        {\cal M}^2_{\tilde{q},LL} & {\cal M}^2_{\tilde{q},LR} \\[2mm]
        {\cal M}^2_{\tilde{q},RL} & {\cal M}^2_{\tilde{q},RR} \end{array} \right),
 \label{EqMassMatrix}
\end{equation}
with $\tilde{q}=\tilde{u},\tilde{d}$. The left-left and right-right blocks in eq.~(\ref{EqMassMatrix}) are given by
\begin{eqnarray}
    & &{\cal M}^2_{\tilde{u},LL} = V_{\rm CKM} M_Q^2 V_{\rm CKM}^{\dag} + D_{\tilde{u},LL}{\bf 1} + \hat{m}^2_u, \nonumber \\
    & &{\cal M}^2_{\tilde{u},RR} = M_U^2 + D_{\tilde{u},RR}{\bf 1} + \hat{m}^2_u, \nonumber \\
    & & {\cal M}^2_{\tilde{d},LL} = M_Q^2 + D_{\tilde{d},LL}{\bf 1} + \hat{m}^2_d,  \nonumber \\
    & & {\cal M}^2_{\tilde{d},RR} = M_D^2 + D_{\tilde{d},RR}{\bf 1} + \hat{m}^2_d,
     \label{EqM2LLRR}
\end{eqnarray}
where $M_{Q,U,D}$ are the hermitian soft supersymmetry (SUSY)-breaking mass matrices of the squarks and
$\hat{m}_{u,d}$ are the diagonal mass matrices of the up-type and down-type quarks.
$D_{\tilde{q},LL}$ and $D_{\tilde{q},RR}$ are the D terms. 
Due to the $SU(2)_{\rm L}$ symmetry the left-left blocks of the up-type and down-type squarks 
in eq.~(\ref{EqM2LLRR}) are related by the CKM matrix $V_{\rm CKM}$.
The left-right and right-left blocks of eq.~(\ref{EqMassMatrix}) are given by
\begin{eqnarray}
 {\cal M}^2_{\tilde{u},RL} = {\cal M}^{2\dag}_{\tilde{u},LR} &=&
\frac{v_2}{\sqrt{2}} T_U - \mu^* \hat{m}_u\cot\beta, \nonumber \\
 {\cal M}^2_{\tilde{d},RL} = {\cal M}^{2\dag}_{\tilde{d},LR} &=&
\frac{v_1}{\sqrt{2}} T_D - \mu^* \hat{m}_d\tan\beta,
\end{eqnarray}
where $T_{U,D}$ are the soft SUSY-breaking trilinear 
coupling matrices of the up-type and down-type squarks entering the Lagrangian 
${\cal L}_{int} \supset -(T_{U\alpha \beta} \ti{u}^\dagger _{R\a}\ti{u}_{L\b}H^0_2 $ 
$+ T_{D\alpha \beta} \ti{d}^\dagger _{R\a}\ti{d}_{L\b}H^0_1)$,
$\mu$ is the higgsino mass parameter, and $\tan\beta$ is the ratio of the vacuum expectation 
values of the neutral Higgs fields $v_2/v_1$, with $v_{1,2}=\sqrt{2} \left\langle H^0_{1,2} \right\rangle$.
The squark mass matrices are diagonalized by the $6\times6$ unitary matrices $U^{\tilde{q}}$,
$\tilde{q}=\tilde{u},\tilde{d}$, such that
\begin{eqnarray}
&&U^{\tilde{q}} {\cal M}^2_{\tilde{q}} (U^{\tilde{q} })^{\dag} = {\rm diag}(m_{\tilde{q}_1}^2,\dots,m_{\tilde{q}_6}^2)\,,
\end{eqnarray}
with $m_{\tilde{q}_1} < \dots < m_{\tilde{q}_6}$.
The physical mass eigenstates $\sq_i, i=1,...,6$ are given by $\sq_i =  U^{\sq}_{i \alpha} \sq_{0\alpha} $.

We define the QFV parameters in the up-type squark sector 
$\delta^{LL}_{\alpha\beta}$, $\delta^{uRR}_{\alpha\beta}$
and $\delta^{uRL}_{\alpha\beta}$ $(\alpha \neq \beta)$ as follows:
\begin{eqnarray}
\delta^{LL}_{\alpha\beta} & \equiv & M^2_{Q \alpha\beta} / \sqrt{M^2_{Q \alpha\alpha} M^2_{Q \beta\beta}}~,
\label{eq:InsLL}\\[3mm]
\delta^{uRR}_{\alpha\beta} &\equiv& M^2_{U \alpha\beta} / \sqrt{M^2_{U \alpha\alpha} M^2_{U \beta\beta}}~,
\label{eq:InsRR}\\[3mm]
\delta^{uRL}_{\alpha\beta} &\equiv& (v_2/\sqrt{2} ) T_{U\alpha \beta} / \sqrt{M^2_{U \alpha\alpha} M^2_{Q \beta\beta}}~,
\label{eq:InsRL}
\end{eqnarray}
where $\alpha,\beta=1,2,3 ~(\alpha \ne \beta)$ denote the quark flavors $u,c,t$.
The relevant mixings in this study are $\ti{c}_R - \ti{t}_L$, 
$\ti{c}_L - \ti{t}_R$, $\ti{c}_R-\ti{t}_R$, and $\ti{c}_L - \ti{t}_L$ 
mixings which are described by the QFV parameters $\delta^{uRL}_{23}$, 
$\delta^{uLR}_{23} \equiv ( \delta^{uRL}_{32})^*$, $\delta^{uRR}_{23}$, and $\dll$, respectively.
We also consider $\ti{t}_L - \ti{t}_R$ mixing described by the quark flavor conserving (QFC) parameter 
$\delta^{uRL}_{33}$ which is defined by eq.~(\ref{eq:InsRL}) with $\alpha= \beta = 3$. 
All QFV parameters and $\delta^{uRL}_{33} $ are assumed to be real.

\section{Reference QFV MSSM scenario}

\begin{figure}[ht]
\centering
\includegraphics[width=48mm]{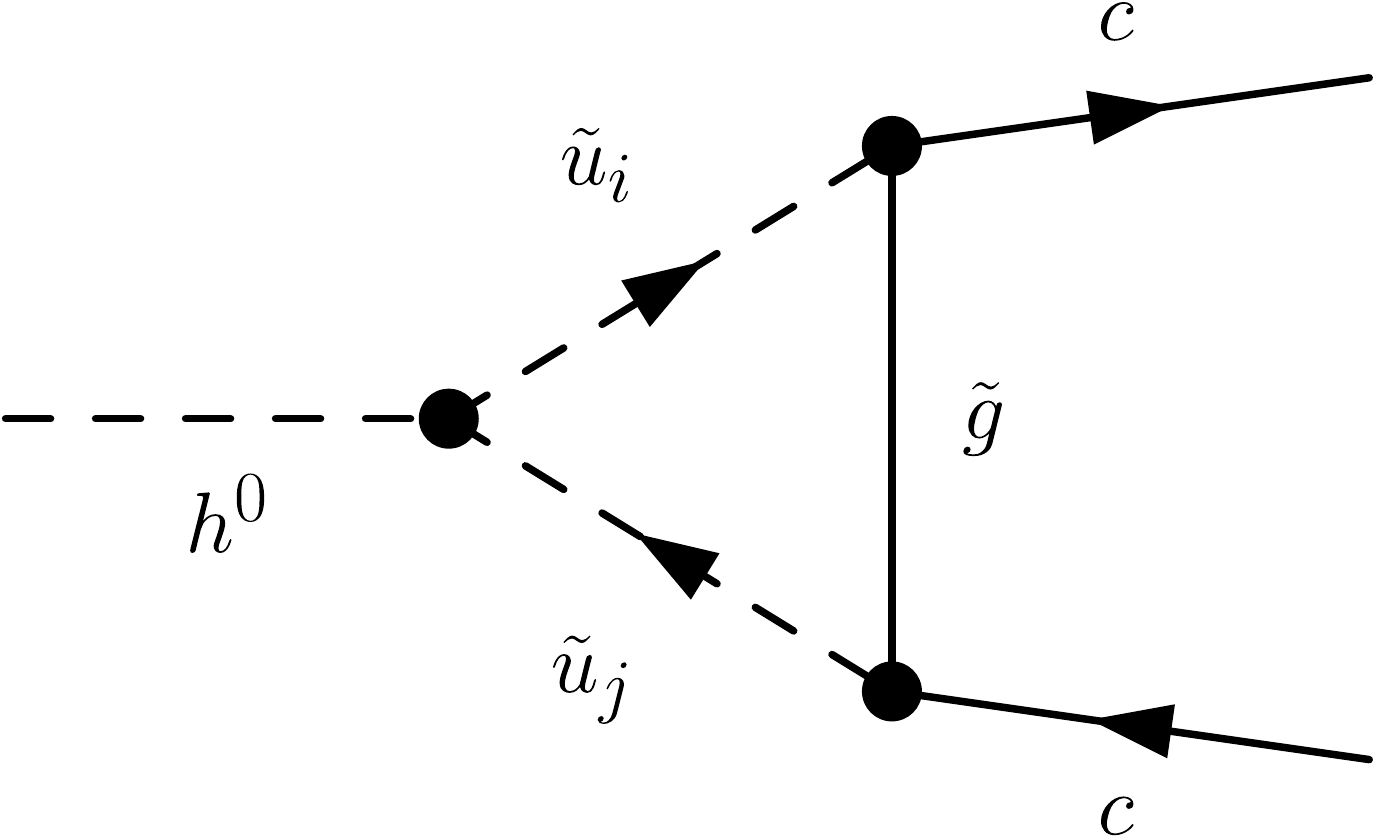}
\caption{Gluino-loop vertex correction to $h^0 \to c \bar{c}$.} \label{gluino_loop}
\end{figure}

%
Our reference scenario of the MSSM with non-minimal QFV is shown in 
Table~\ref{basicparam} \cite{Bartl}. Table~\ref{physmasses} shows the resulting 
physical masses of the involved particles. Table~\ref{flavordecomp} shows the 
flavor decomposition of the lightest up-type squarks $\su_1$ and $\su_2$. 
The reference scenario contains large $\ti c_{L/R}-\ti t_{L/R}$ mixings and large QFV trilinear 
couplings of $\ti c_L-\ti t_R-H_2^0$ and $\ti c_R-\ti t_L-H_2^0$, where $\ti c_{L/R}$ 
and $\ti t_{L/R}$ are the SUSY partners of the charm and top quark, respectively. 
It satisfies (i) the constraint from the observed Higgs boson mass at LHC, (ii) the recent LHC limits 
on the masses of squarks, gluino, charginos and neutralinos, (iii) the strong constraints on QFV 
from the B-meson experiments, such as BR($b \to s \gamma$), BR($B_s \to \mu^+ \mu^-$) 
and $\Delta M_{B_s}$, (iv) the constraints on the trilinear couplings from the vacuum stability 
conditions and (v) the experimental limit on SUSY contributions to the electroweak $\rho$  
parameter. In this decoupling Higgs scenario, $h^0 \simeq Re(H_2^0)$, the lightest 
up-type squarks $\ti u_1$  and $\ti u_2$ are strong mixtures of $\ti c_{L/R}-\ti t_{L/R}$, and the trilinear 
couplings ($\ti c_L-\ti t_R-h^0$, $\ti c_R-\ti t_L-h^0$, $\ti t_L-\ti t_R-h^0$ couplings) are large; 
therefore, $\ti u_{1,2}-\ti u_{1,2}-h^0$ couplings are large. This results in an enhancement of the 
$\ti u_{1,2}-\ti u_{1,2}-\ti g$ loop vertex correction to the decay amplitude of $h^0 \to c \bar{c}$ 
shown in Fig.~\ref{gluino_loop}, where $\ti g$ is a gluino. Thus, this can lead to a large deviation 
of the MSSM prediction for the decay width $\Gamma (h^0 \to c \bar{c})$ from the SM prediction.

\begin{table}
\begin{center}
\begin{tabular}{|c|c|c|}
  \hline
 $M_1$ & $M_2$ & $M_3$ \\
 \hline \hline
 250~\gev  &  500~\gev &  1500~\gev \\
  \hline
\end{tabular}
\vskip0.4cm
\begin{tabular}{|c|c|c|}
  \hline
 $\mu$ & $\tan \beta$ & $m_{A^0}$ \\
 \hline \hline
 2000~\gev & 20 &  1500~\gev \\
  \hline
\end{tabular}
\vskip0.4cm
\begin{tabular}{|c|c|c|c|}
  \hline
   & $\alpha = 1$ & $\alpha= 2$ & $\alpha = 3$ \\
  \hline \hline
   $M_{Q \alpha \alpha}^2$ & $(2400)^2~\gev^2$ &  $(2360)^2~\gev^2$  & $(1850)^2~\gev^2$ \\
   \hline
   $M_{U \alpha \alpha}^2$ & $(2380)^2~\gev^2$ & $(1050)^2~\gev^2$ & $(950)^2~\gev^2$ \\
   \hline
   $M_{D \alpha \alpha}^2$ & $(2380)^2~\gev^2$ & $(2340)^2~\gev^2$ &  $(2300)^2~\gev^2$  \\
   \hline
\end{tabular}
\vskip0.4cm
\begin{tabular}{|c|c|c|c|}
  \hline
   $\delta^{LL}_{23}$ & $\delta^{uRR}_{23}$  &  $\delta^{uRL}_{23}$ & $\delta^{uLR}_{23}$\\
  \hline \hline
   0.05 & 0.2 &  0.03   &  0.06  \\
    \hline
\end{tabular}
\end{center}
\caption{Reference QFV scenario: shown are the basic MSSM parameters 
at $Q = 125.5~{\rm GeV} \simeq m_{h^0}$,
except for $m_{A^0}$ which is the pole mass (i.e.\ the physical mass) of $A^0$, 
with $T_{U33} = - 2050$~GeV (corresponding to $\delta^{uRL}_{33} = - 0.2$). All other squark 
parameters not shown here are zero. $M_{1,2,3}$ are the U(1), SU(2), SU(3) gaugino mass parameters.}
\label{basicparam}
\end{table}
%
\begin{table}
\begin{center}
\begin{tabular}{|c|c|c|c|c|c|}
  \hline
  $\mnt{1}$ & $\mnt{2}$ & $\mnt{3}$ & $\mnt{4}$ & $\mch{1}$ & $\mch{2}$ \\
  \hline \hline
  $260$ & $534$ & $2020$ & $2021$ & $534$ & $2022$ \\
  \hline
\end{tabular}
\vskip 0.4cm
\begin{tabular}{|c|c|c|c|c|}
  \hline
  $m_{h^0}$ & $m_{H^0}$ & $m_{A^0}$ & $m_{H^+}$ \\
  \hline \hline
  $126.08$  & $1498$ & $1500$ & $ 1501$ \\
  \hline
\end{tabular}
\vskip 0.4cm
\begin{tabular}{|c|c|c|c|c|c|c|}
  \hline
  $\msg$ & $\msu{1}$ & $\msu{2}$ & $\msu{3}$ & $\msu{4}$ & $\msu{5}$ & $\msu{6}$ \\
  \hline \hline
  $1473$ & $756$ & $965$ & $1800$ & $2298$ & $2301$ & $2332$ \\
  \hline
\end{tabular}
\end{center}
\caption{
Physical masses in GeV of the particles for the scenario of Table 1. 
}
\label{physmasses}
\end{table}
%
\begin{table}
\begin{center}
\begin{tabular}{|c|c|c|c|c|c|c|c|}
  \hline
  & $\su_L$ & $\sca_L$ & $\sto_L$ & $\su_R$ & $\sca_R$ & $\sto_R$ \\
  \hline \hline
 $\su_1$  & $0$ & $0.0004$ & $0.012$ & $0$ & $0.519$ & $0.468$ \\
  \hline 
  $\su_2$  & $0$ & $0.0004$ & $0.009$ & $0$ & $0.480$ & $0.509$ \\
  \hline
\end{tabular}
\end{center}
\caption{
Flavor decomposition of $\su_1$ and $\su_2$ for the scenario of Table 1. Shown are the squared coefficients.
 }
\label{flavordecomp}
\end{table}

\section{$\Gamma(h^0 \to c \bar{c})$ at full one-loop level in the MSSM with QFV}

We calculate the decay width $\Gamma(h^0 \to c \bar{c})$ at full one-loop level in the general MSSM 
with non-minimal QFV. We adopt the $\drbar$ renormalization scheme taking the renormalization scale 
as $Q = 125.5~{\rm GeV} \simeq m_{h^0}$. 
In the general MSSM at one-loop level, in addition to the diagrams that contribute 
within the SM, $\Gamma(h^0 \to c \bar{c})$ also receives contributions from loop 
diagrams with additional Higgs bosons and supersymmetric particles. The flavor violation 
is induced by one-loop diagrams with squarks that have a mixed quark flavor nature. 
The one-loop contributions to $\Gamma(h^0 \to c \bar{c})$ contain three parts, 
QCD ($g$) corrections, SUSY-QCD ($\ti{g}$) corrections and electroweak corrections 
including loops with neutralinos, charginos and Higgs bosons.  
Details of the computation of the decay width at full one-loop level are described in Ref.~\cite{Bartl}. 

\section{Numerical results}

%



\begin{figure}[ht]
\centering
\includegraphics[width=70mm]{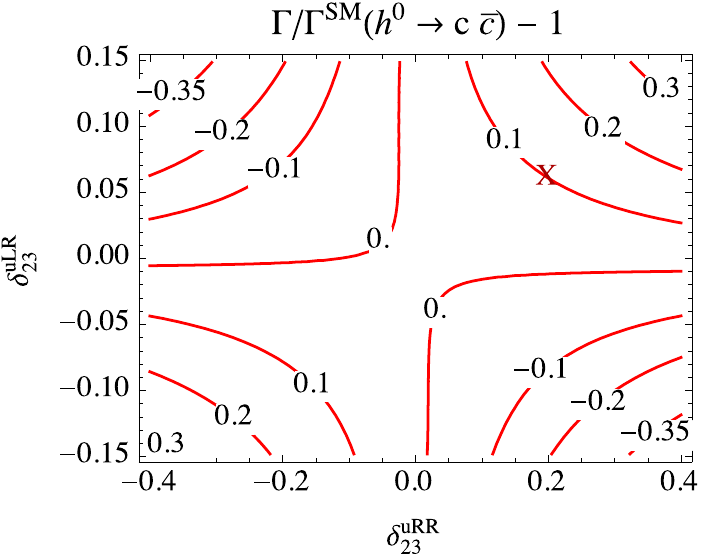}
\caption{
Contour plot of the deviation of the full one-loop level MSSM width $\Gamma (h^0 \to c \bar{c})$ 
from the SM width $\Gamma^{SM} (h^0 \to c \bar{c})$ for our reference QFV scenario of Table 1. 
The shown range satisfies all the relevant experimental and theoretical constraints described in Section 3. 
}
\label{contour plot of deviation}
\end{figure}

Fig.~\ref{contour plot of deviation} shows the contour plot of the deviation of the MSSM prediction from 
the SM prediction $\Gamma^{SM} (h^0 \to c \bar{c})$ = 0.118 MeV \cite{Gamma_SM} in the $\durr$-$\dulr$ 
plane, where $\durr$ and $\dulr$ are the $\ti c_R-\ti t_R$ and $\ti c_L-\ti t_R$ mixing parameters, respectively. 
As can be seen, the MSSM prediction is very sensitive to the QFV parameters $\durr$ and $\dulr$, and 
the deviation from the SM prediction can be very large (as large as $\sim \pm 35\%$). 
We have found that the MSSM prediction becomes nearly equal to the SM prediction 
if we switch off all the QFV parameters in our reference QFV scenario.

\section{Observability of the deviation at ILC}

\begin{figure}[ht]
\centering
\includegraphics[width=70mm]{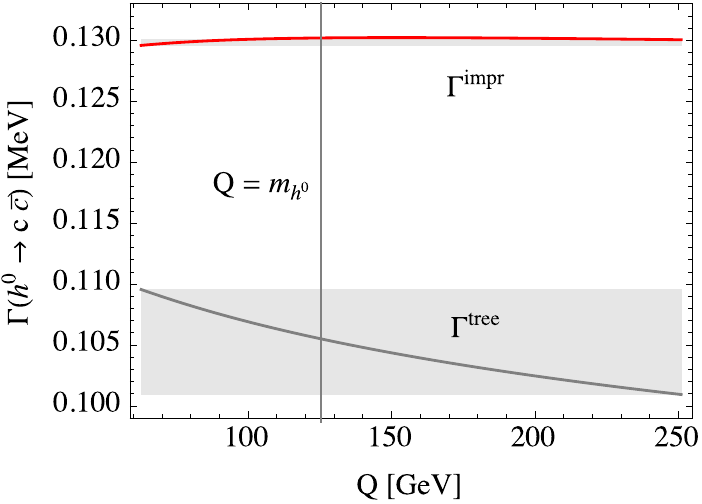}
\caption{Renormalization-scale dependence of $\Gamma(h^0 \to c \bar{c})$ for the reference scenario of Table 1.  $\Gamma^{\rm impr}(h^0 \to c \bar{c})$ is the improved one-loop corrected width. The vertical line shows ${\rm Q}=m_{h^0}$. }
\label{qdep}
\end{figure}

The observation of any significant deviation of the decay width $\Gamma (h^0 \to c \bar{c})$ from its SM 
prediction implies 'New Physics' beyond the SM. It is very important to estimate the theoretical and 
experimental uncertainties of the width reliably in order to confirm such a deviation. The relative theoretical error 
of the SM width is estimated to be $\sim 6\%$ \cite{Almeida}. The relative theoretical error of the MSSM width 
is also estimated to be $\sim 6\%$ \cite{Bartl}. The theoretical error stems from the parametric uncertainties 
(such as uncertainties of the charm quark mass $m_c$ and the QCD coupling $\alpha_s$) and the 
scale-dependence uncertainty. The latter uncertainty is estimated by varying the renormalization scale 
Q from ${\rm Q}/2$ to $2{\rm Q}$. (Note that in our case ${\rm Q}=m_{h^0}$.) 
The uncertainty due to the renormalization scale Q dependences of the MSSM width is 
small ($\simeq 0.5\%$) as can be seen in Fig.\ref{qdep}. 
As shown in Fig.\ref{contour plot of deviation}, the deviation 
of the MSSM width from the SM width can be as large as $\sim \pm 35\%$. Such a large deviation can be 
observed at a future $e^+ e^-$ collider ILC (International Linear Collider) with a CM energy 500 GeV 
and an integrated luminosity of 1600 $fb^{-1}$, where the expected experimental error of the width is 
$\sim 3\%$ \cite{Tian}. A measurement of the width at LHC is a hard task because of the difficulties 
in charm-tagging.

\section{Conclusion}

We have calculated the width $\Gamma (h^0 \to c \bar{c})$  at full one-loop level within the MSSM 
with non-minimal  quark flavor violation. We have studied $\ti{c}_{R,L}-\ti{t}_{R,L}$ mixings, taking into 
account the experimental constraints from B-meson data, the Higgs mass measurement and SUSY 
particle searches. 
We have found the width $\Gamma (h^0 \to c \bar{c})$ is very sensitive to the QFV parameters, in particular to 
$\ti{c}_{R,L}-\ti{t}_{R,L}$ mixings. Whereas in the QFC MSSM case $\Gamma (h^0 \to c \bar{c})$ is 
only slightly different from its SM value, in the QFV case it can deviate from the SM value by up to 
$\sim \pm 35\%$. After taking into account the experimental and theoretical uncertainties of the decay width, 
we conclude that these QFV SUSY effects can be observed at ILC. Therefore, we have a good opportunity 
to discover the QFV SUSY effect in this decay $h^0 \to c \bar{c}$ at ILC. 

%
\acknowledgments{
This work is supported by the "Fonds zur F\"orderung der
wissenschaftlichen Forschung (FWF)" of Austria, project No. P26338-N27.
}


\end{document}